\documentclass{elsart}

\usepackage{epsfig}

\usepackage{amssymb}

\begin{document}

\begin{frontmatter}



\title{On pricing of interest rate derivatives}


\author[Canberra]{T. Di Matteo},
\author[Milano]{M. Airoldi} 
and
\author[Alessandria,cor1]{E. Scalas}
\corauth[cor1]{Corresponding author: Tel: +39 0131 283854, fax: +39 0131 283841.}
\ead{scalas@cicladi.unipmn.it}
\address[Canberra]{Department of Applied Mathematics \\
Research School of Physical Sciences and Engineering \\
The Australian National University
Canberra ACT 0200, Australia.}
\address[Milano]{Financial Engineering, Mediobanca S.p.A., \\
Piazzetta Enrico Cuccia 1, Milan, Italy.}
\address[Alessandria]{Dipartimento di Scienze e Tecnologie Avanzate, 
			Universit\`a del Piemonte Orientale, 
			Corso Borsalino 54, 
			I--15100 Alessandria, Italy.}

\begin{abstract}
% Text of abstract
At present, there is an explosion of practical interest in the pricing of interest rate (IR) derivatives. Textbook pricing methods do not take into account the leptokurticity of the underlying IR process. In this paper, such a leptokurtic behaviour is illustrated using LIBOR data, and a possible martingale pricing scheme is discussed.
\end{abstract}


\end{frontmatter}

% main text
\section{Introduction}
\label{sec1}
In financial theory and practice, interest rates are a very
important subject which can be approached from several different
perspectives. The classical theoretical approach models the term structure of
interest rates using stochastic processes. Various models have been proposed and can be found in \cite{Pagan,Rebonato,Rebonato03}.
Although they provide analytical formulas for the pricing of interest rate derivatives, the implied deformations of the term
structure have a Brownian motion component and are often rejected
by empirical data (see \cite{Chan}). The inadequacies of the Gaussian
model for the description of financial time series has been
reported since a long time ago by Mandelbrot \cite{MandelbrotB}, but thanks
to the availability of large sets of financial data, the interest
on this point has risen recently \cite{LibrMant,LibDac}.
In particular, the fat-tail property of the empirical distribution
of price changes has been widely documented and is a crucial
feature for monitoring the extreme risks and for accurately pricing interest rate derivatives. An important recent development in the pricing of interest rate derivatives is the emergence of models that incorporate lognormal
volatilities for forward LIBOR or forward swap rates while keeping
interest rates stable \cite{Zhao}. To our knowledge, up to now, no universally accepted theory
has been obtained for the description of interest rates data \cite{Nuyts}.

In this framework, we have empirically studied the probability
density distribution of LIBOR, in order to characterize the
stochastic behavior of the daily fluctuations. In Section \ref{sec2}, we present the data set and the data
analysis. Section \ref{sec3} contains a short discussion of a
possible IR derivative pricing scheme using martingale methods. 
\begin{figure}
\begin{center}
\begin{tabular}{cc}
\mbox{\epsfig{file=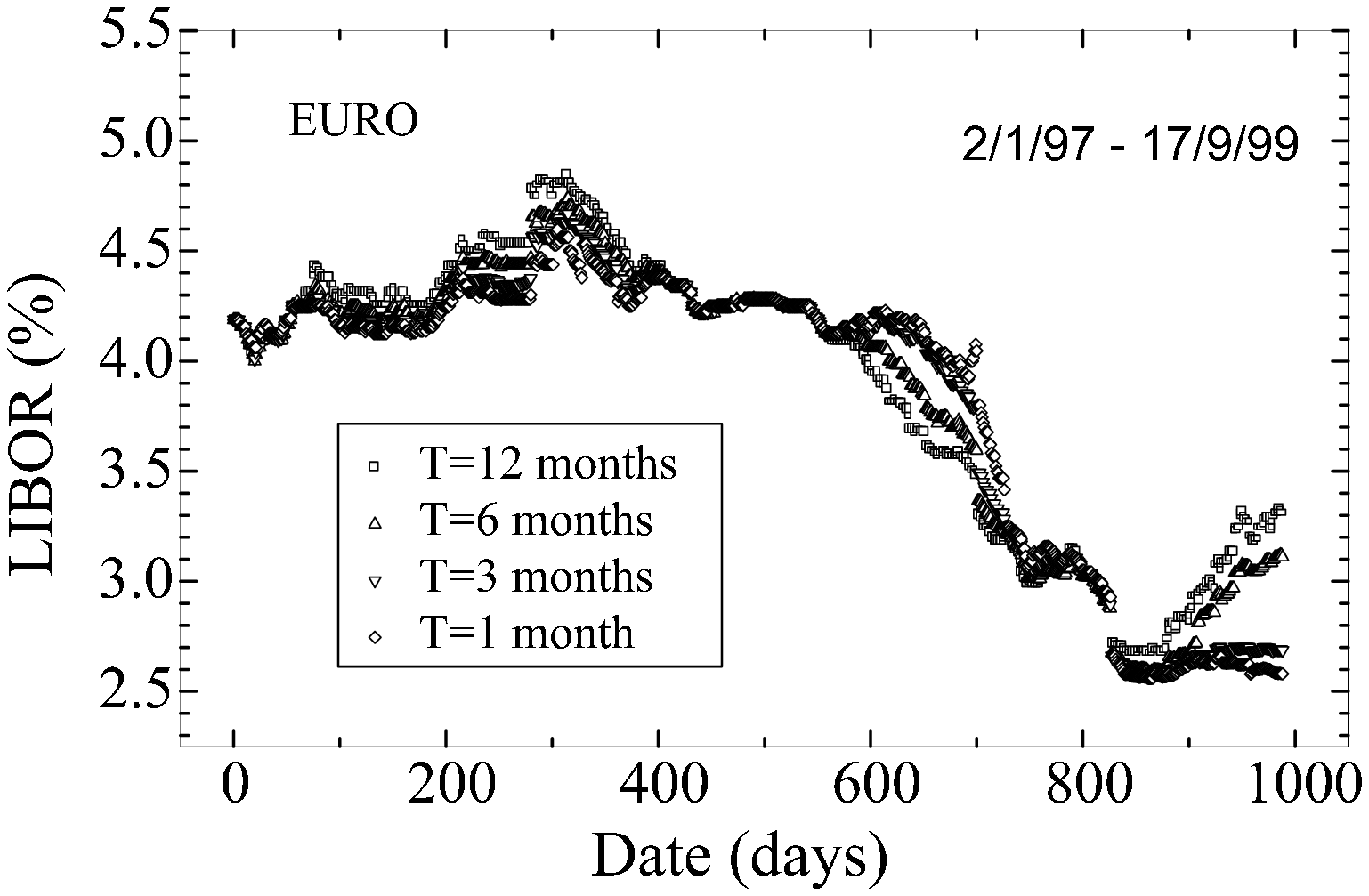,width=7.cm,angle=0}}
&\mbox{\epsfig{file=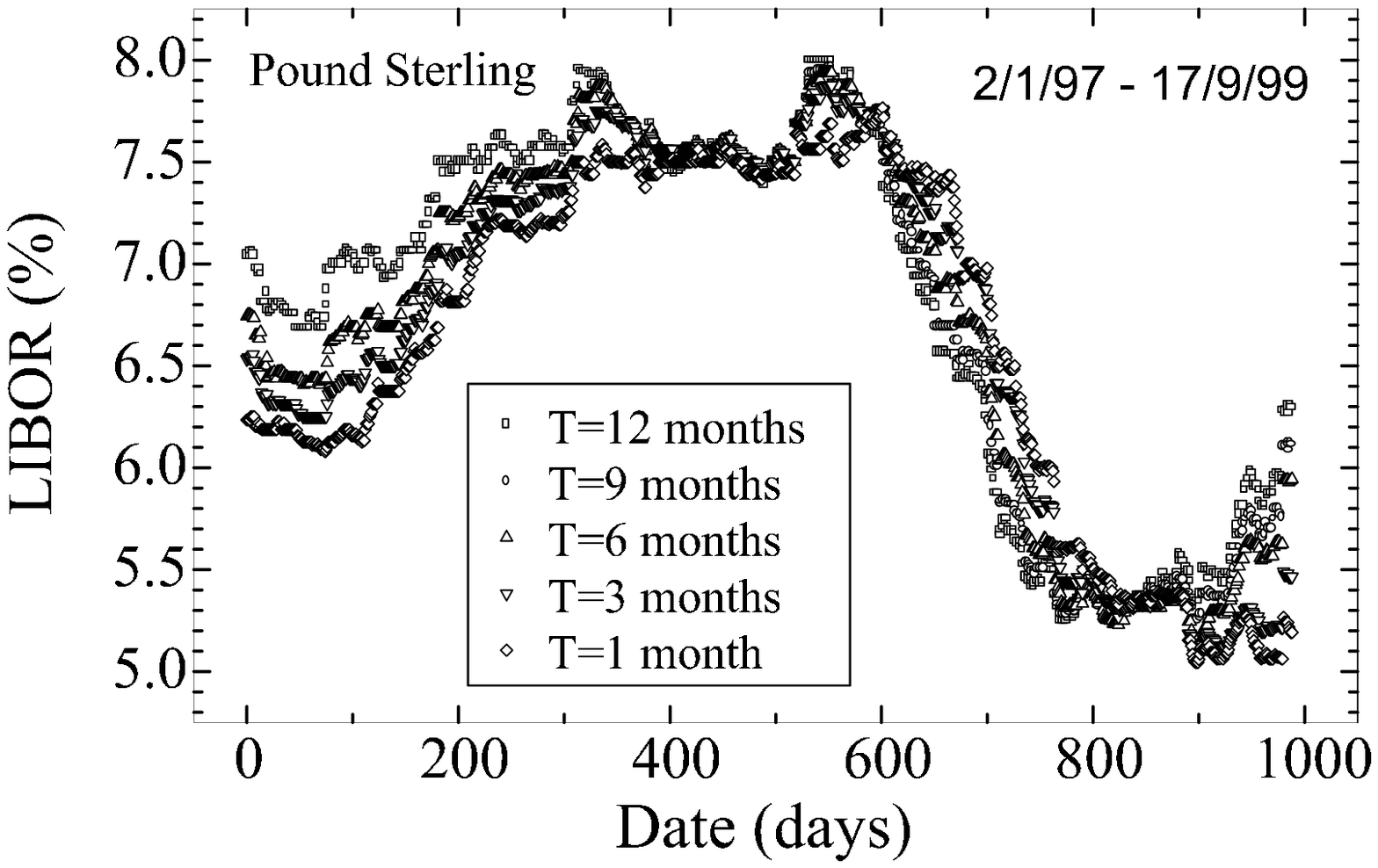,width=7.cm,angle=0}}
\end{tabular}
\end{center}
\caption{LIBOR interest rates $r(T,t)$ as a function of the
current date $t$, for the maturities $T$=$1$, $3$, $6$ and $12$
months and EURO currency (Left) and Pound Sterling currency (Right).}
\label{f.1}
\end{figure}

\section{Empirical findings}
\label{sec2}

\begin{figure}
\begin{center}
\begin{tabular}{cc}
\mbox{\epsfig{file=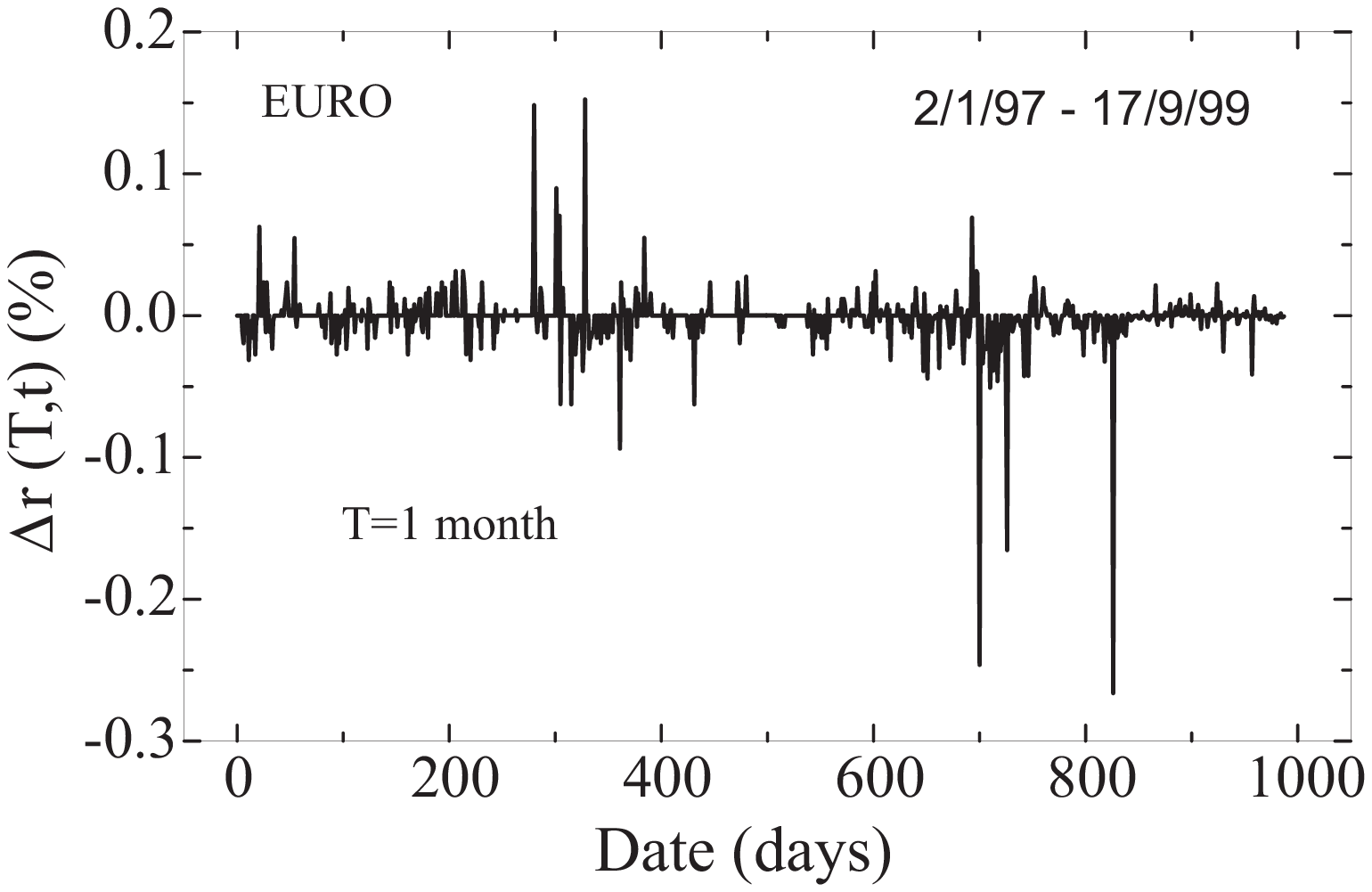,width=7.cm,angle=0}}
&\mbox{\epsfig{file=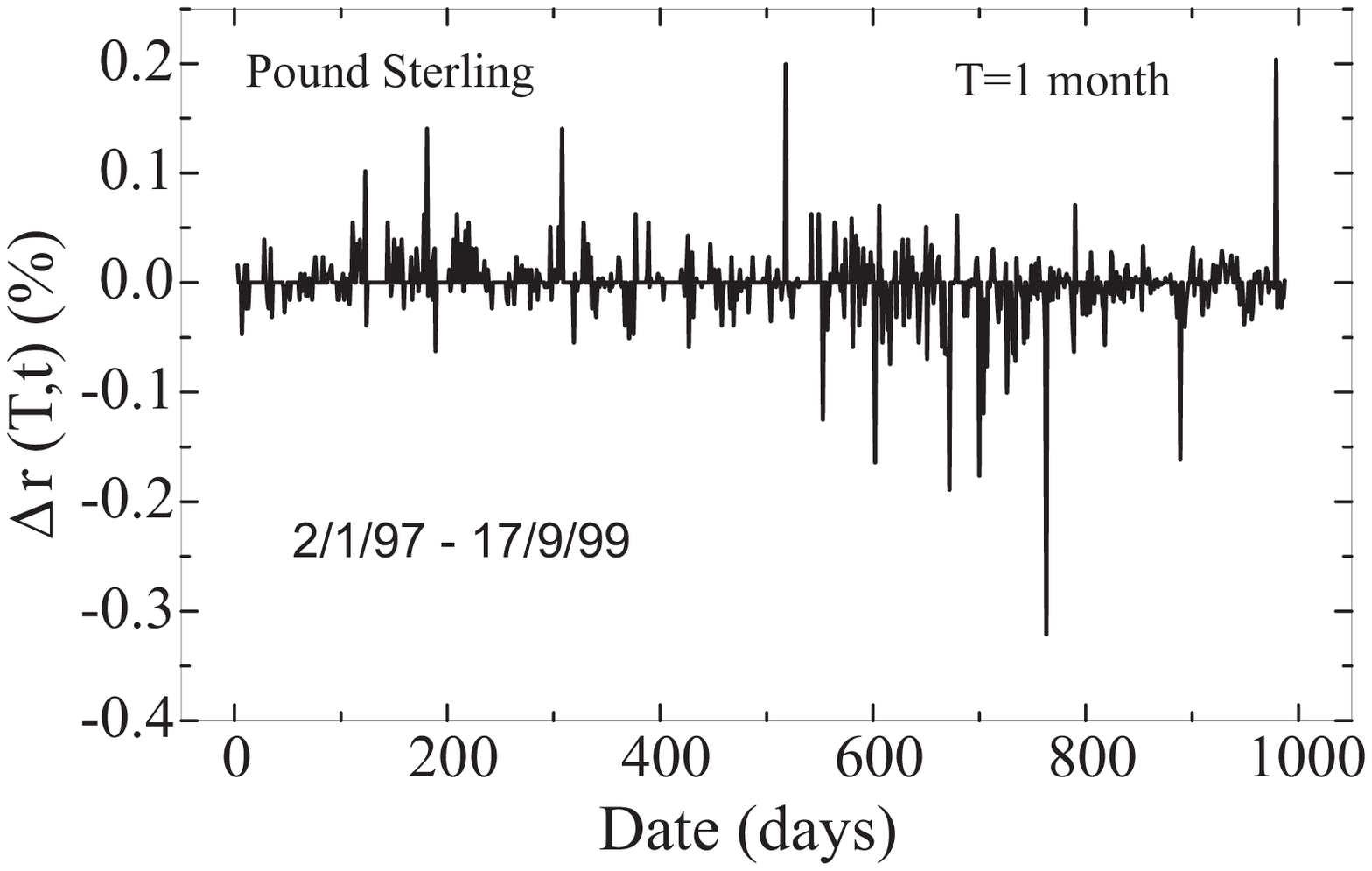,width=7.cm,angle=0}}
\end{tabular}
\end{center}
\caption{1-month LIBOR increments as a function of the current date for EURO (Left) and for Sterling Pound (Right).}
\label{f.3}
\end{figure}

\begin{figure}
\begin{center}
\mbox{\epsfig{file=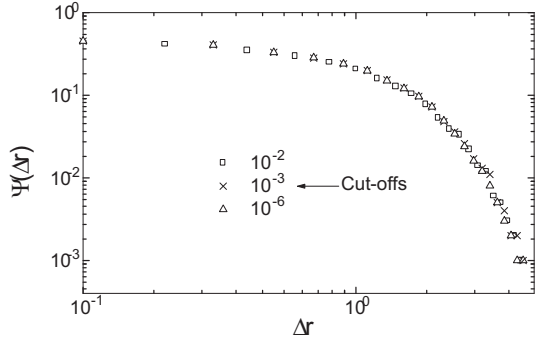,width=7.cm,angle=0}}
\end{center}
\caption{The complementary cumulative distribution function of a
simulated Gaussian stochastic process using different decimal digit
precisions (cut-offs). }
\label{f.4}
\end{figure}

\begin{figure}
\begin{center}
\begin{tabular}{cc}
\mbox{\epsfig{file=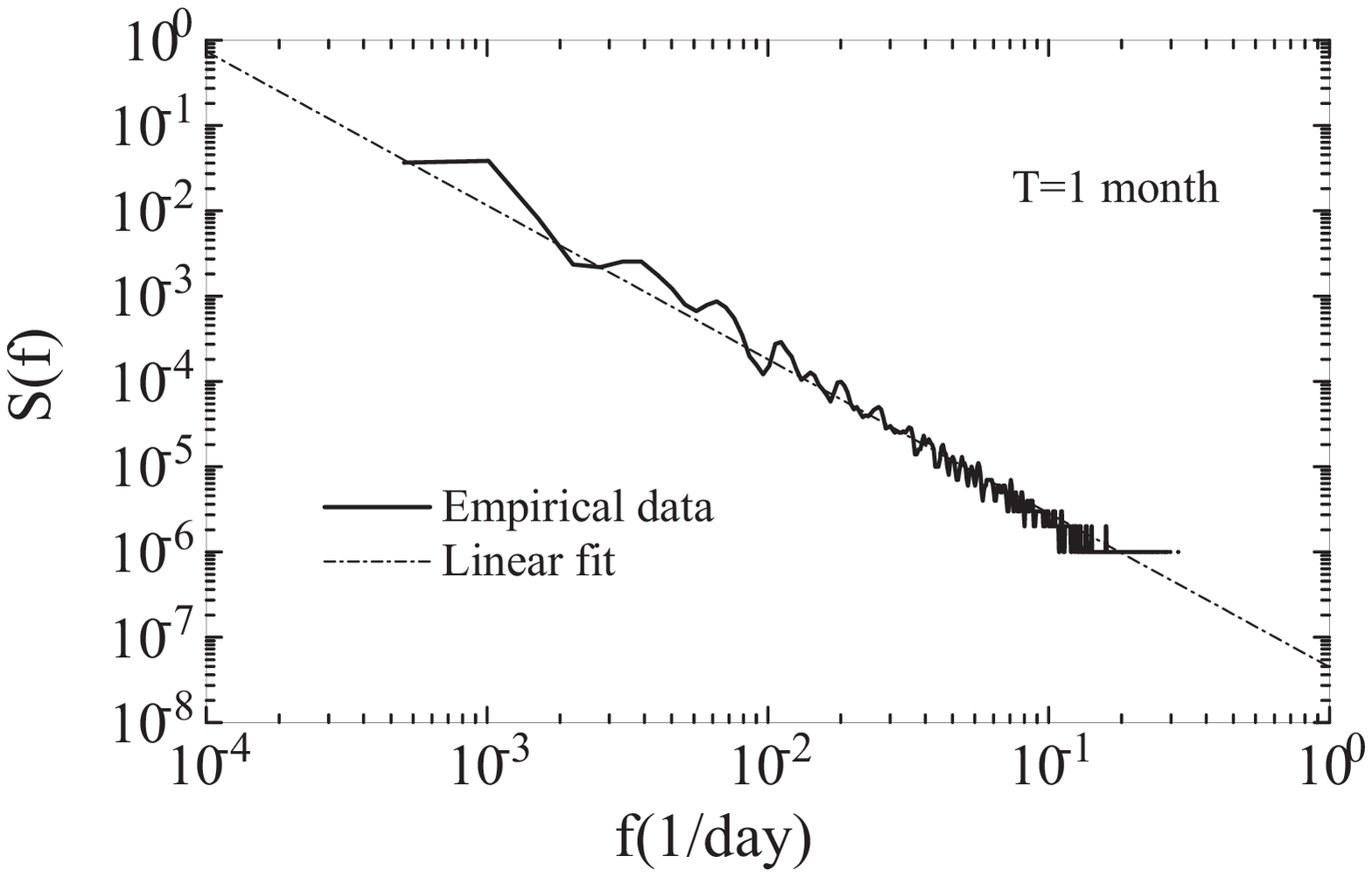,width=7.cm,angle=0}}
&\mbox{\epsfig{file=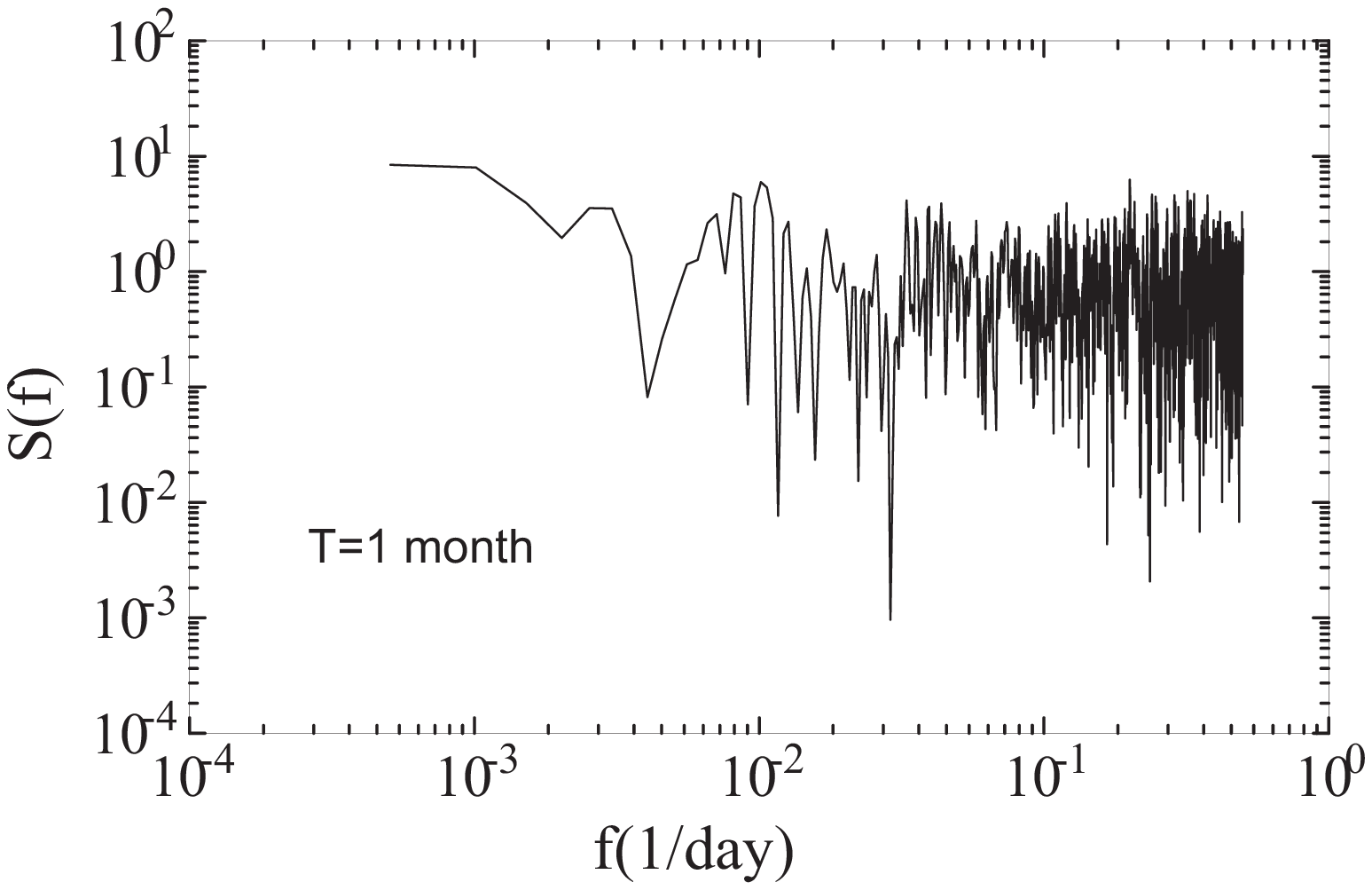,width=7.cm,angle=0}}
\end{tabular}
\end{center}
\caption{Power spectrum of $r(T,t)$ (Left) and of $\Delta r(T,t)$ (Right) for the EURO. $T$=1 month and $\Delta t$=1 day.}
\label{f.6}
\end{figure}

\begin{figure}
\begin{center}
\begin{tabular}{cc}
\mbox{\epsfig{file=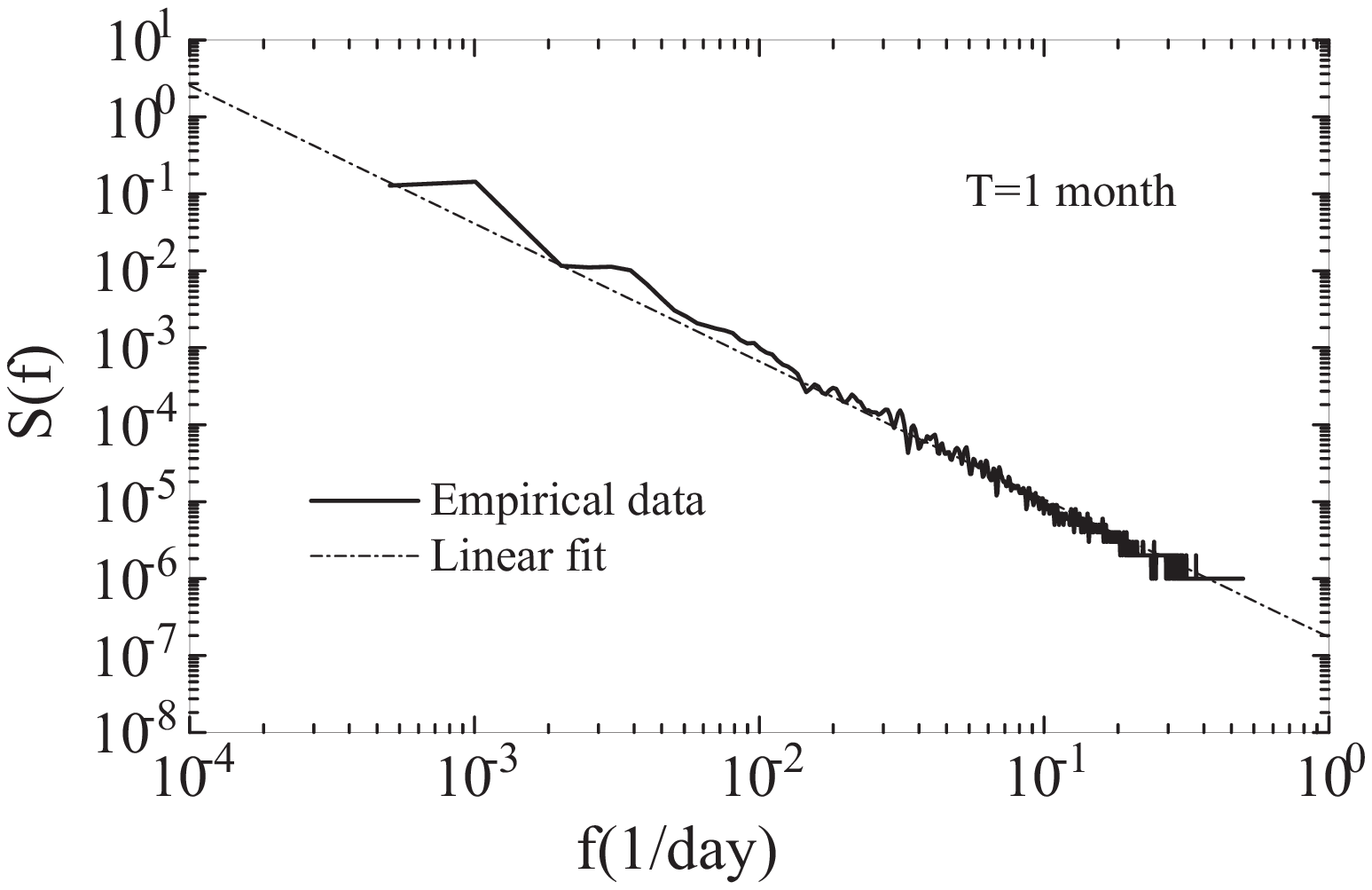,width=7.cm,angle=0}}
&\mbox{\epsfig{file=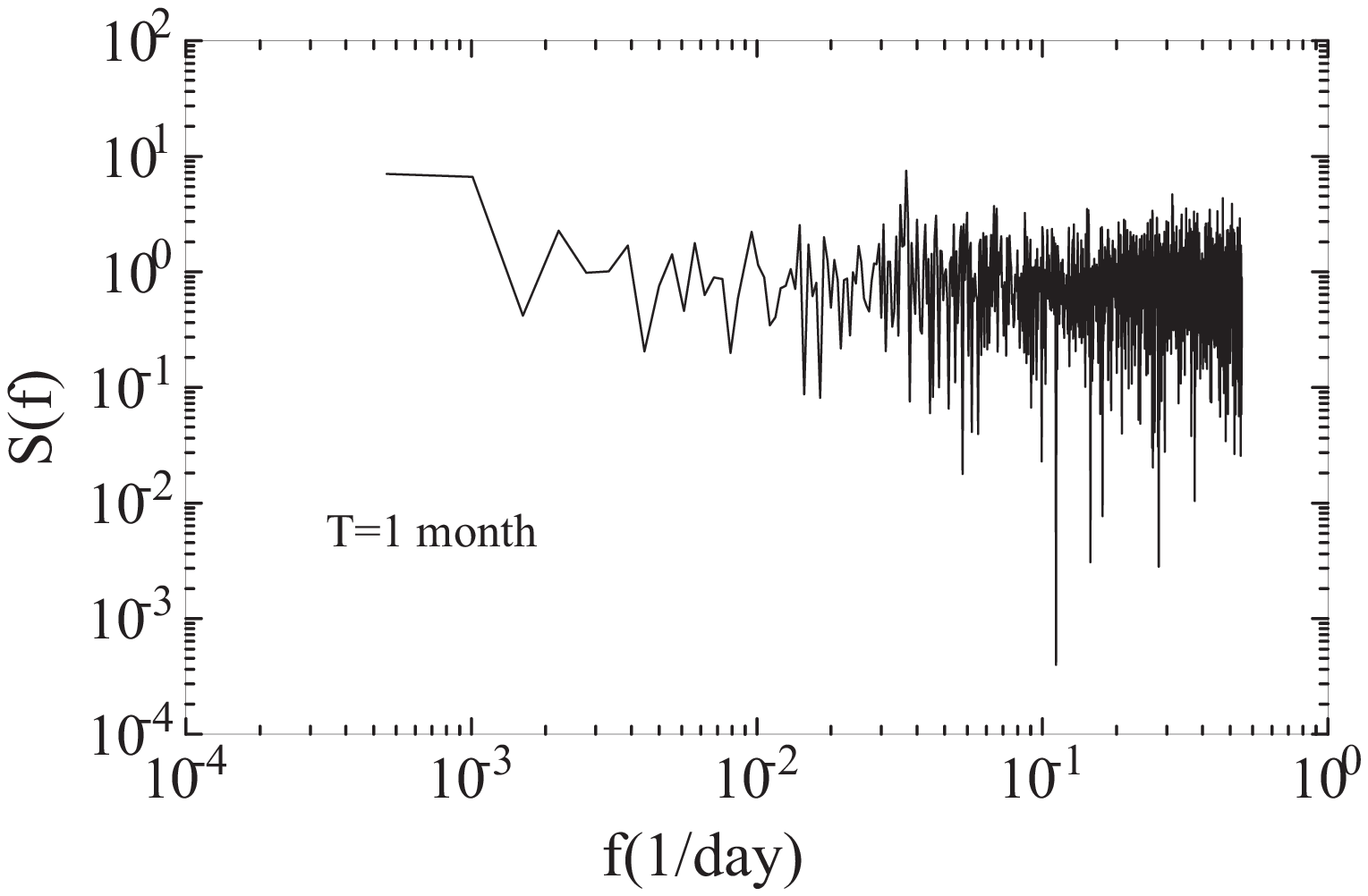,width=7.cm,angle=0}}
\end{tabular}
\end{center}
\caption{Power spectrum of $r(T,t)$ (Left) and of $\Delta r(T,t)$ (Right) for the Sterling Pound. $T$=1 month and $\Delta t$=1 day.}
\label{f.7}
\end{figure}

LIBOR stands for the London Interbank Offered Rate and is the rate
of interest at which banks are willing to offer deposits to other
prime banks, in marketable size, in the London interbank market. BBA (British Bankers' Association) LIBOR \cite{bba} is the most
widely used benchmark or reference rate. It is used as the basis for settlement of interest rate
contracts on many of the world's major future and option exchanges
as well as most Over the Counter and lending transactions. 
\begin{figure}
\begin{center}
\begin{tabular}{cc}
\mbox{\epsfig{file=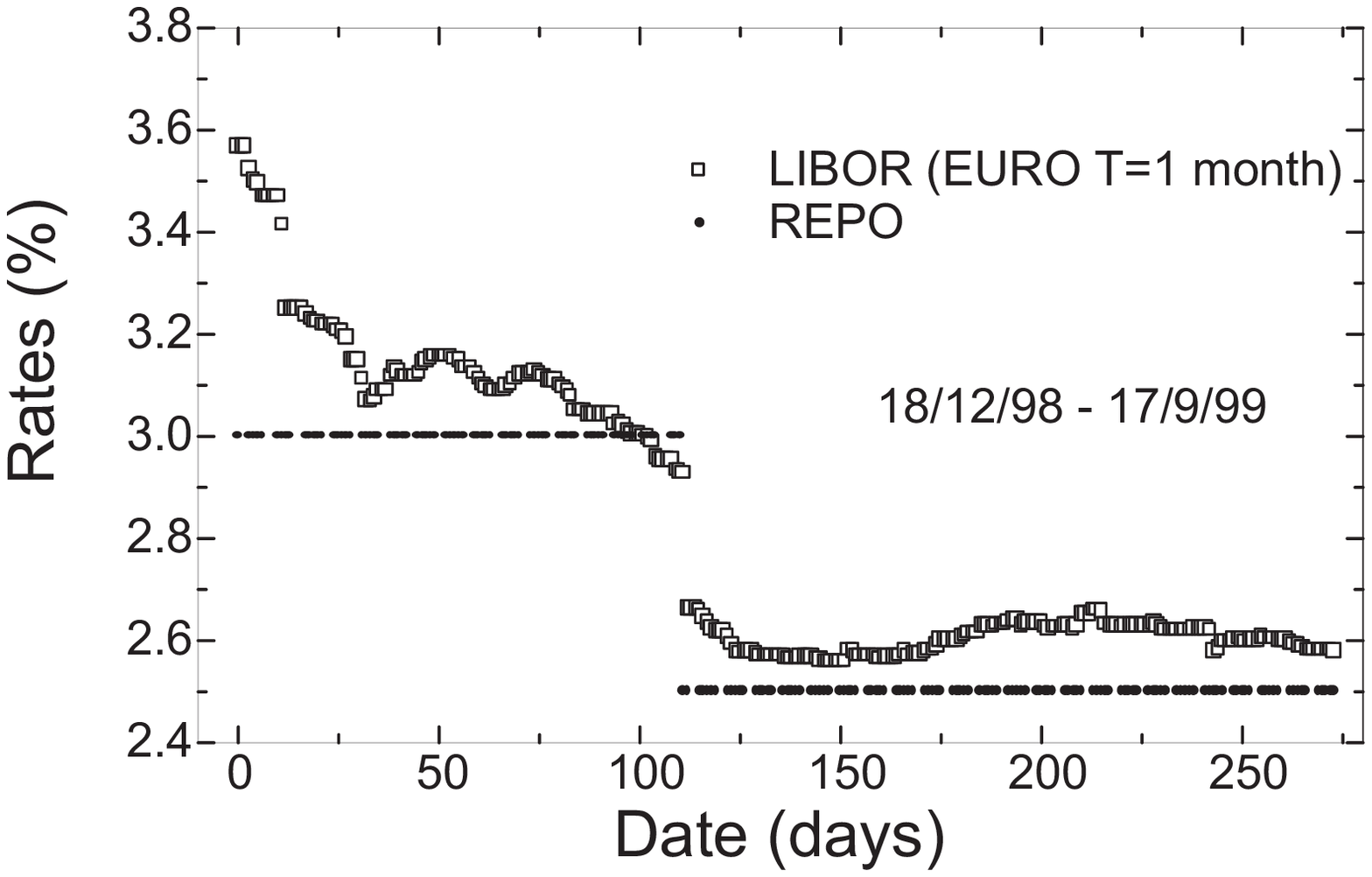,width=7.cm,angle=0}}
&\mbox{\epsfig{file=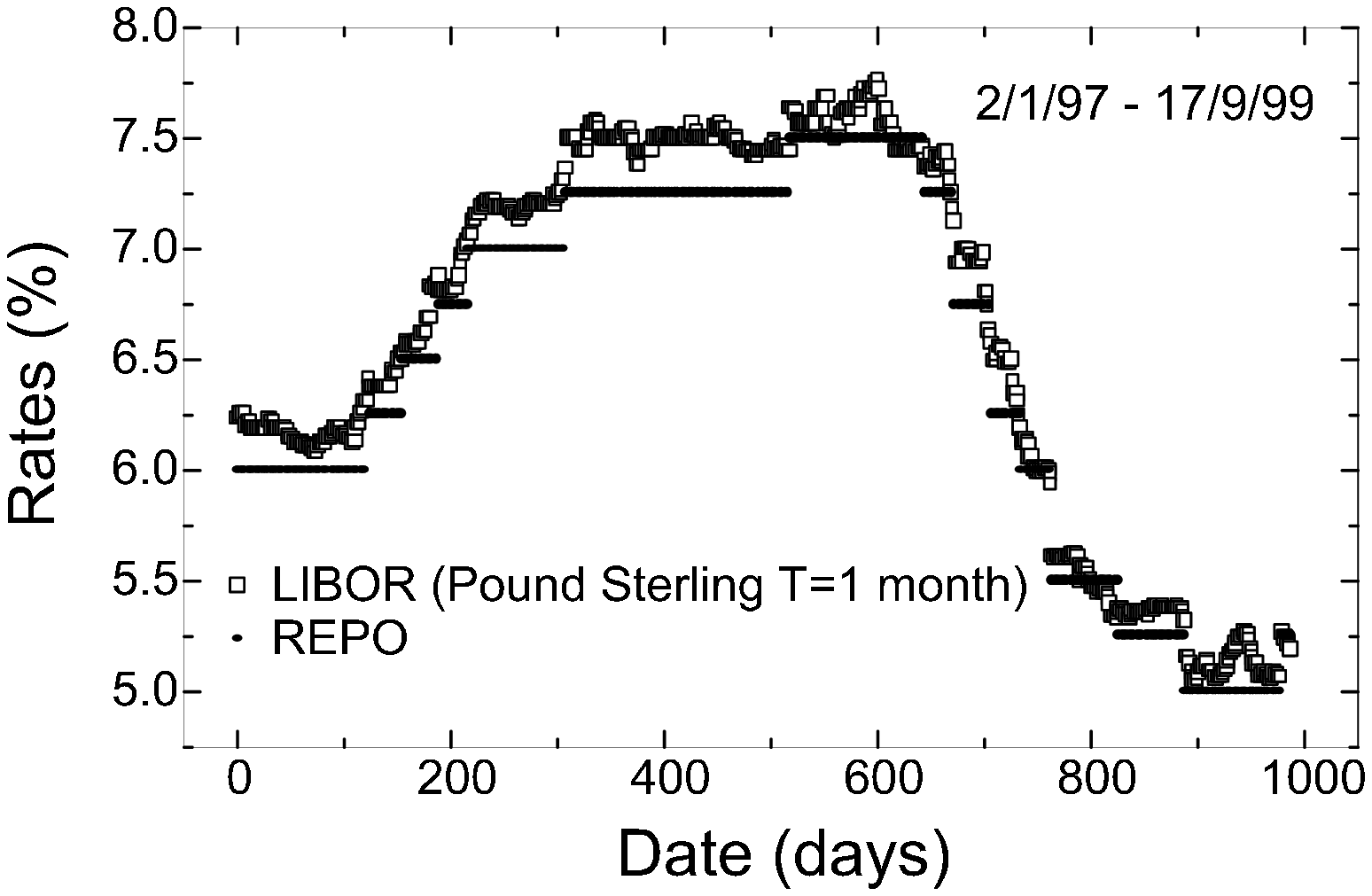,width=7.cm,angle=0}}
\end{tabular}
\end{center}
\caption{LIBOR compared to REPO for EURO (Left) and for Sterling Pound (Right).}
\label{f.2}
\end{figure}
BBA LIBOR is compiled each working day and broadcast through ten
international distribution networks. BBA LIBOR fixings are provided in seven international currencies:
Pound Sterling, US Dollar, Japanese Yen, Swiss Franc, Canadian
Dollar, Australian Dollar, EURO. LIBOR rates are fixed for each
currency at monthly maturities from one month to $12$ months.
Rates shall be contributed in decimal to at least two decimal
places but no more than five. In the following, we have analyzed a data set of LIBOR interest
rates $r(T,t)$, where $T$ is the maturity date and $t$ the current
date, for EURO and Pound Sterling. These data are shown in
Fig.\ref{f.1} where $t$ goes from January 2,
$1997$ to September $17$, $1999$, and $T$ assumes the following
values: $1$, $3$, $6$, $9$ and $12$ months for the Pound Sterling
and $1$, $3$, $6$, $12$ months for the EURO. In Fig.\ref{f.2} the 1-month LIBOR is compared to the
interest rates fixed by Central Banks at that time, namely the
REPO (repurchase agreement) data. It is quite evident that BBA LIBOR follows the trend
determined by the decisions of Central Banks. In order to roughly
eliminate these trends, in Fig.\ref{f.3} the
interest rates differences $\Delta r(T,t)$=$r(T,t+\Delta
t)-r(T,t)$, with $\Delta t$ being $1$ day and $T$=$1$ month, are
plotted as a function of the current date, for the EURO and the
Pound Sterling, respectively. A similar behavior is also found for the other maturities. Some large oscillations of $\Delta
r$ are induced by Central Banks. In any case, $\Delta r$ heavily
fluctuates around zero. We focus the attention on the probability
distribution behavior of the interest rates increments $\Delta
r(T,t)$. 
To this purpose, we
estimate $\Psi (\Delta r)$, the complementary cumulative
distribution function of the daily interest rates increments,
defined as:
\begin{equation}
\label{uno} \Psi (\Delta r)=1-\int_{-\infty}^{\Delta r} p(\eta)
d\eta
\end{equation}
\noindent 
where $p$ is the probability density of $\Delta r(T,t)$. Because LIBOR data are supplied with only few decimal digits, it
is interesting to examine the effects of different data cut-offs
in the behavior of $\Psi (\Delta r)$. In Fig.\ref{f.4}, we plot
the complementary cumulative distribution function for a simulated Gaussian
stochastic process using data characterized by three different
decimal digit precisions. 
\begin{figure}
\begin{center}
\begin{tabular}{cc}
\mbox{\epsfig{file=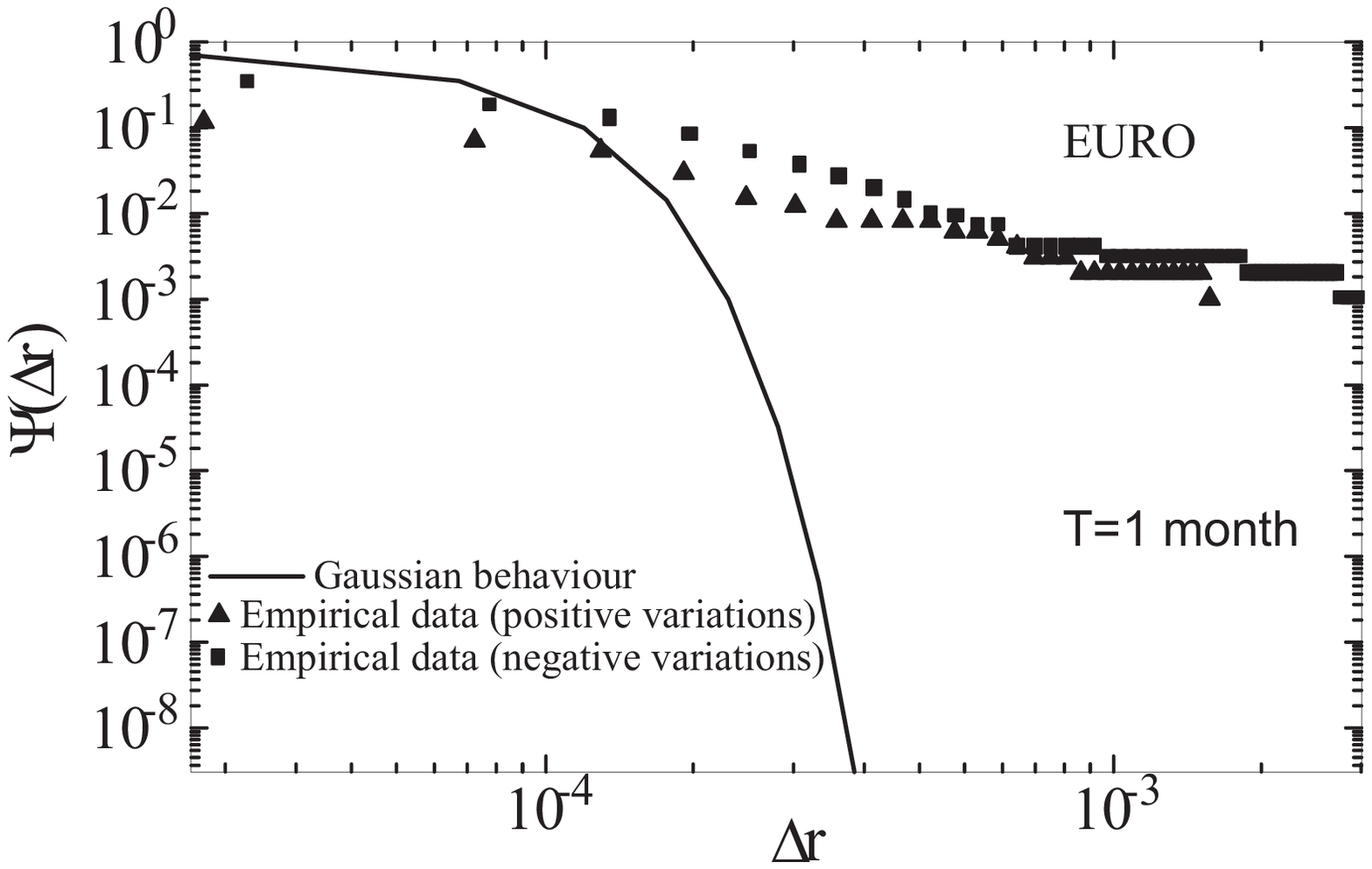,width=7.cm,angle=0}}
&\mbox{\epsfig{file=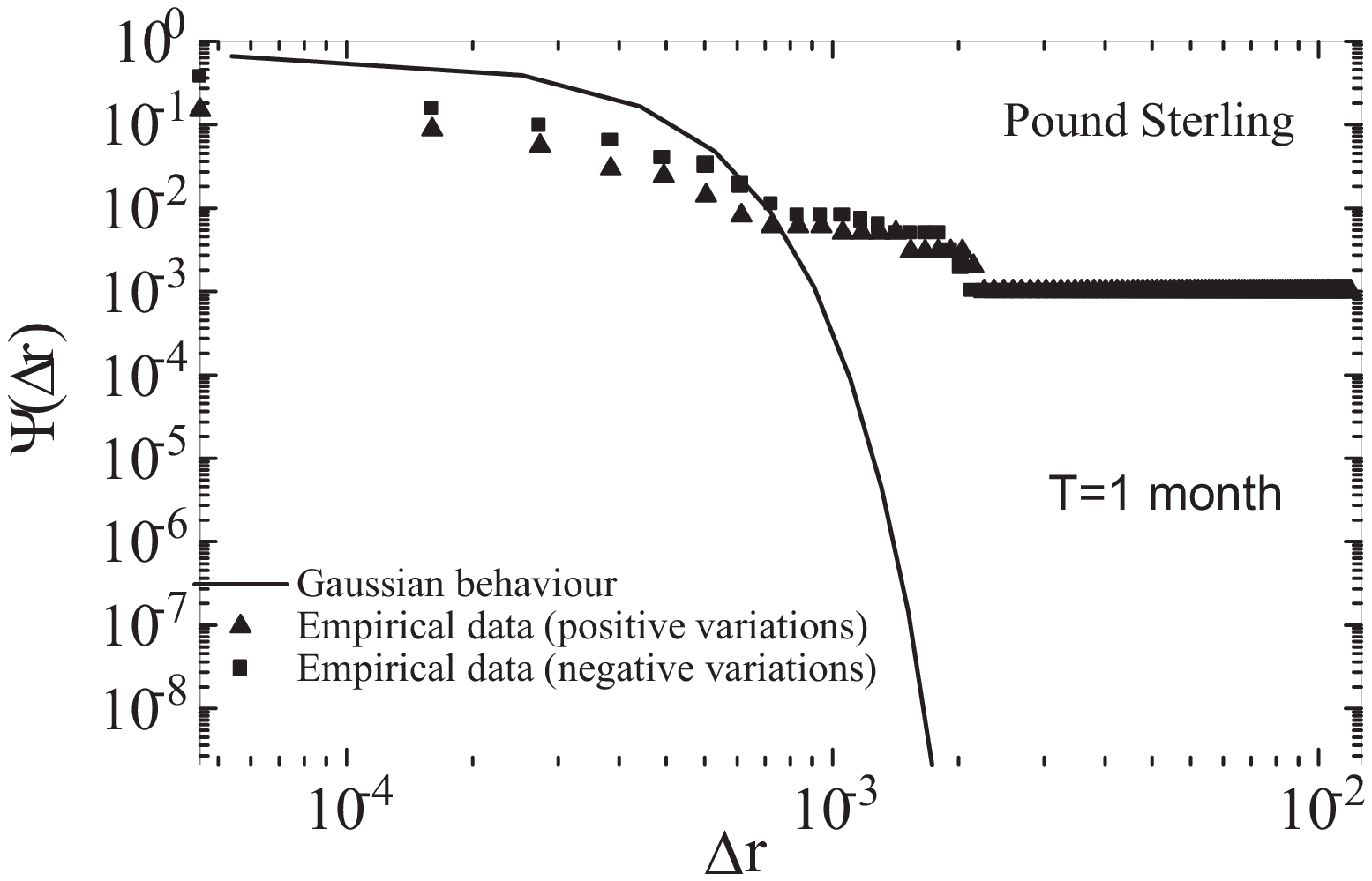,width=7.cm,angle=0}}
\end{tabular}
\end{center}
\caption{The complementary cumulative distribution function of
LIBOR increments for EURO (Left) and Sterling Pound (Right). $T$=1 month and $\Delta t$=1 day.}
\label{f.5}
\end{figure}
It turns out that the numerical rounding
does not influence the results.
Fig.\ref{f.5} shows the tail distribution behaviors
in the case of EURO and Sterling Pound, respectively. In
particular, in Fig.\ref{f.5} (Left side), we report the empirical results
obtained estimating the probability density function of both
positive and negative LIBOR increments with $\Delta t$=$1$ day and
$T$=$1$ month. These empirical curves are slightly asymmetric and
the negative variations are more probable than the positive one.
In the same figure, these two curves are compared with the
equivalent (i.e. with the same average and standard deviation)
Gaussian complementary cumulative distribution. The non-Gaussian
behavior is also evident from Fig.\ref{f.5} (Right side). In both
Figures \ref{f.5}, the empirical LIBOR data exhibit a
{\it fat tail} or {\it leptokurtic} character, which is present for the other maturities as well. These observations
indicate that the random behavior of $\Delta r(T,t)$ is
non-Gaussian and that using a Gaussian probability density
function leads to underestimating the probability of large
fluctuations. For a better understanding of the deviations from a pure Brownian motion and what kind of stochastic process we are dealing with, we analyze the power spectral
density behavior. The power spectra \cite{Kay}, $S(f)$, for both $r(1,t)$ and $\Delta r(1,t)$
are reported in Fig.\ref{f.6} for the EURO and in
Fig.\ref{f.7} for the Sterling Pound. For $r$ the
spectral density shows a power law behavior. A linear fit gives a
slope value $\alpha={-1.80 \pm 0.02}$ and $\alpha={-1.79 \pm
0.01}$ for EURO and Sterling Pound, respectively. A similar result
holds for the other maturities. Therefore, we argue that the power
spectrum analysis for $r(T,t)$ indicates a stochastic process with
spectral components decreasing as $S(f) \sim {f}^{\alpha}$
\cite{Feller}. The power spectrum for the increments (Figs.\ref{f.6} and
\ref{f.7} (Right)) is flat, typical of a white noise
process. These results are also corroborated by
a similar analysis
performed on Eurodollars interest rates for a longer time period \cite{DiMatteo}.

\section{Discussion}
\label{sec3}

In the previous section, we have shown that the daily increment of
the interest rate series is non-Gaussian, non-Brownian and follows a
leptokurtic distribution. The problem arises of how derivatives
written on interest rates can be evaluated, given that the usual
Gaussian white-noise assumption of many models is not satisfied.
Indeed, a first partial answer, is that the central limit theorem
ensures that, after a sufficiently long time, the increment
distribution will tend to a Gaussian distribution. However, if the
time horizon of derivative evaluation is not appropriate,
deviations from the Gaussian behavior may lead to a dramatic
underestimate of large increments with a consequent improper risk
coverage as well as option price estimate. Although well studied in mainstream
finance, this problem has received much attention in recent times, within
the community of physicists working on financial problems. In particular, Bouchaud and
Sornette \cite{Sornette 1994} suggested the direct use of the
historical probability measure, rather than the equivalent
martingale measure for evaluating options. In this way, one gets
an option price depending on the expected rate of returns, a
consequence which is not fully desirable due to the subjective
character of that rate. Assessing trends is a difficult
task, as they depend on decision taken by Central Banks (as shown
in Fig.\ref{f.2}) and are based on macroeconomic
effects. Thus, martingale methods could prove more
reliable. As early as 1977, some years after the seminal paper of
Black and Scholes, Parkinson generalized their approach to option
pricing and explicitly took into account leptokurtic distributions \cite{Parkinson}. More recently, Boyarchenko and Levendorskii have
studied the problem of option pricing in the presence of a
specific distribution which seems to fit well the empirical data
in many instances: the truncated L\'evy distribution
\cite{Boyarchenko}. Wim Schoutens has recently
published a book on L\'evy processes in finance devoted to the
extension of martingale methods to a large class of leptokurtic 
distributions \cite{Schoutens03}. 

The method can be described from a heuristic point of view. Let
$S(t)$ denote the stochastic process underlying a contingent claim $C(S,t)$; thus, $S(t)$ can
be a price process, an interest rate process, etc..
Let $X(t) = \log S(t)$ be the corresponding logarithmic process. Let further
$p(x,t)$ be the probability density of finding the value $x$ of the random
variable $X$ at time $t$ (this is a conditional probability density in $x$ with respect to
suitable initial conditions; here, $X(0)=0$). This density defines the probability measure ${\bf P}$.
It is possible to show that, if the following relation holds true:
\begin{equation}
\label{exponential}
E_{{\bf P}} \{\exp[a X(t)] \} = \int_{-\infty}^{+\infty} \exp(ax) \; p(x,t) \; dx = \exp[g(a) t]
\end{equation}
where $E_{{\bf P}} $ denotes the expectation operator with respect to ${\bf P}$, $a$ is a complex number and $g(a)$ a complex
function of $a$, then the process $\xi(t,a) = \exp [ aX(t)-g(a) ]$ is a martingale with
respect to the measure ${\bf P}$. Therefore, as a consequence of Girsanov's theorem,
we can build an equivalent martingale measure ${\bf Q}_{T,a}$ such that the 
Radon-Nikodym derivative $d {\bf Q}_{T,a} / d {\bf P}$ is given by:
\begin{equation}
\label{radon}
\frac{d {\bf Q}_{T,a}}{d {\bf P}} = \xi (T,a).
\end{equation}
In order to price a contingent claim written on $S(t) = S(0) \exp[X(t)]$, we require that
the discounted process $S_d (t)$ is a martingale with respect to the measure ${\bf Q}_{T,a}$.
This is equivalent to the requirement that the process $\xi(t,a) S_d (t)$ is a martingale with respect to 
${\bf P}$. In this paper, we have shown that interest rates are fluctuating variables. However, just
for the sake of simplicity, let us consider a fixed interest rate $r$. In this particular case, the martingale condition
is equivalent to the following equation:
\begin{equation}
\label{a}
g(a+1) - g(a) - r = 0.
\end{equation}
In other words, if it is possible to determine a single value of $a$ such that Eq. (\ref{a}) is satisfied,
the measure ${\bf Q}_{T,a}$ exists and is unique. For instance, if $S(t)$ is described by geometric Brownian 
motion with drift $\mu$ and volatility $\sigma$, we have $g(a) = \mu a - (i a \sigma)^{2}/2$ and there is a unique solution of Eq. (\ref{a}): $a = - (\mu+\sigma^{2}/2-r)/\sigma^{2}$. The reader is referred to \cite{Boyarchenko} for the case of truncated L\'evy processes. Then, under the requirement that $C(S,t)$ be a martingale with respect to the measure ${\bf Q}_{T,a}$, one can find the price of the contingent claim. If $C_{d} (S,t)$ is the discounted process,
we have:
\begin{equation}
\label{optionprice}
C_{d} (S,t) = \xi^{-1}(t,a) E_{{\bf P}}[ \xi(T,a) C_{d}(S,T)|F_t]
\end{equation}
where $F_t$ is the appropriate filtration.

The technique has been already applied to derivatives written on interest rates, here, we outline the generalization of a popular IR model: the Heath, Jarrow and Morton (HJM) model, following Eberlein and Reible \cite{Eberlein99}.
Within this model, the zero-coupon bond price $P(T,t)$ is given by:
\begin{equation}
\label{eq2}
P(T,t) = P(T,0) \exp \left[ \int_0^t r(s,s)\;ds
\right] \frac{\exp[\int_0^t \sigma(T,s)\;dW_s]}{E\{\exp[\int_0^t
\sigma(T,s)\;dW_s]\}},
\end{equation}
where $\sigma(T,t)$ is the bond volatility structure, $W_t$ is the Wiener process, $E$ is the expectation operator and $r(s) := r(s,s)$. Note that $r(T,t)$ has been interpreted as the instantaneous forward rate $f(T,t)$. The Wiener process can be replaced by a leptokurtic L\'evy process $L_t$, such that the expectation in the denominator is finite. In this case, it is possible to show that the discounted bond-price process is a martingale and that the martingale measure is unique \cite{Raible00}. Also, the European vanilla call option price on a bond maturing at time $T$ can be obtained \cite{Eberlein99}. 
Along these lines, we believe, it is
possible to develop a consistent option pricing theory taking
into account the leptokurtic character of the empirical short to
mid-term interest rate distributions. 

\section*{Acknowledgements}
We thank the Risk Management division of Banca Intesa (Milano, Italy) for
providing us the data. This work was partially supported by grants from the Italian
M.I.U.R. Project COFIN 2003 ``Order and Chaos in nonlinear extended
systems: coherent structures, weak stochasticity and anomalous transport'' and by the Italian M.I.U.R. 
F.I.S.R. Project ``Ultra-high frequency dynamics of financial markets''.

\end{document}